\documentclass[aps,prl,preprint,superscriptaddress,showpacs]{revtex4}
\usepackage{graphicx}
\usepackage{wrapfig}
\usepackage{multirow}
\usepackage[usenames]{color}

\begin{document}



\title{Production and $\beta$ decay of rp-process nuclei $^{96}$Cd, $^{98}$In and $^{100}$Sn}

\author{D. Bazin}
\email[]{bazin@nscl.msu.edu}
\affiliation{National Superconducting Cyclotron Laboratory, Michigan State University, East Lansing, MI 48824, USA}

\author{F. Montes}
\affiliation{National Superconducting Cyclotron Laboratory, Michigan State University, East Lansing, MI 48824, USA}
\affiliation{Joint Institute for Nuclear Astrophysics, Michigan State University, East Lansing, MI 48824, USA}

\author{A. Becerril}
\affiliation{National Superconducting Cyclotron Laboratory, Michigan State University, East Lansing, MI 48824, USA}
\affiliation{Department of Physics and Astronomy, Michigan State University, East Lansing, MI 48824, USA}

\author{G. Lorusso}
\affiliation{National Superconducting Cyclotron Laboratory, Michigan State University, East Lansing, MI 48824, USA}
\affiliation{Department of Physics and Astronomy, Michigan State University, East Lansing, MI 48824, USA}

\author{A. Amthor}
\affiliation{National Superconducting Cyclotron Laboratory, Michigan State University, East Lansing, MI 48824, USA}
\affiliation{Department of Physics and Astronomy, Michigan State University, East Lansing, MI 48824, USA}

\author{T. Baumann}
\affiliation{National Superconducting Cyclotron Laboratory, Michigan State University, East Lansing, MI 48824, USA}

\author{H. Crawford}
\affiliation{National Superconducting Cyclotron Laboratory, Michigan State University, East Lansing, MI 48824, USA}
\affiliation{Department of Chemistry, Michigan State University, East Lansing, MI 48824, USA}

\author{A. Estrade}
\affiliation{National Superconducting Cyclotron Laboratory, Michigan State University, East Lansing, MI 48824, USA}
\affiliation{Department of Physics and Astronomy, Michigan State University, East Lansing, MI 48824, USA}

\author{A. Gade}
\affiliation{National Superconducting Cyclotron Laboratory, Michigan State University, East Lansing, MI 48824, USA}
\affiliation{Department of Physics and Astronomy, Michigan State University, East Lansing, MI 48824, USA}

\author{T. Ginter}
\affiliation{National Superconducting Cyclotron Laboratory, Michigan State University, East Lansing, MI 48824, USA}

\author{C. J. Guess}
\affiliation{National Superconducting Cyclotron Laboratory, Michigan State University, East Lansing, MI 48824, USA}
\affiliation{Joint Institute for Nuclear Astrophysics, Michigan State University, East Lansing, MI 48824, USA}
\affiliation{Department of Physics and Astronomy, Michigan State University, East Lansing, MI 48824, USA}

\author{M. Hausmann}
\affiliation{National Superconducting Cyclotron Laboratory, Michigan State University, East Lansing, MI 48824, USA}

\author{G. W. Hitt}
\affiliation{National Superconducting Cyclotron Laboratory, Michigan State University, East Lansing, MI 48824, USA}
\affiliation{Joint Institute for Nuclear Astrophysics, Michigan State University, East Lansing, MI 48824, USA}
\affiliation{Department of Physics and Astronomy, Michigan State University, East Lansing, MI 48824, USA}

\author{P. Mantica}
\affiliation{National Superconducting Cyclotron Laboratory, Michigan State University, East Lansing, MI 48824, USA}
\affiliation{Department of Chemistry, Michigan State University, East Lansing, MI 48824, USA}

\author{M. Matos}
\affiliation{National Superconducting Cyclotron Laboratory, Michigan State University, East Lansing, MI 48824, USA}
\affiliation{Joint Institute for Nuclear Astrophysics, Michigan State University, East Lansing, MI 48824, USA}

\author{R. Meharchand}
\affiliation{National Superconducting Cyclotron Laboratory, Michigan State University, East Lansing, MI 48824, USA}
\affiliation{Joint Institute for Nuclear Astrophysics, Michigan State University, East Lansing, MI 48824, USA}
\affiliation{Department of Physics and Astronomy, Michigan State University, East Lansing, MI 48824, USA}

\author{K. Minamisono}
\affiliation{National Superconducting Cyclotron Laboratory, Michigan State University, East Lansing, MI 48824, USA}

\author{G. Perdikakis}
\affiliation{National Superconducting Cyclotron Laboratory, Michigan State University, East Lansing, MI 48824, USA}
\affiliation{Joint Institute for Nuclear Astrophysics, Michigan State University, East Lansing, MI 48824, USA}

\author{J. Pereira}
\affiliation{National Superconducting Cyclotron Laboratory, Michigan State University, East Lansing, MI 48824, USA}

\author{J. Pinter}
\affiliation{National Superconducting Cyclotron Laboratory, Michigan State University, East Lansing, MI 48824, USA}
\affiliation{Department of Chemistry, Michigan State University, East Lansing, MI 48824, USA}

\author{M. Portillo}
\affiliation{National Superconducting Cyclotron Laboratory, Michigan State University, East Lansing, MI 48824, USA}

\author{H. Schatz}
\affiliation{National Superconducting Cyclotron Laboratory, Michigan State University, East Lansing, MI 48824, USA}
\affiliation{Joint Institute for Nuclear Astrophysics, Michigan State University, East Lansing, MI 48824, USA}

\author{K. Smith}
\affiliation{National Superconducting Cyclotron Laboratory, Michigan State University, East Lansing, MI 48824, USA}
\affiliation{Joint Institute for Nuclear Astrophysics, Michigan State University, East Lansing, MI 48824, USA}

\author{J. Stoker}
\affiliation{National Superconducting Cyclotron Laboratory, Michigan State University, East Lansing, MI 48824, USA}
\affiliation{Department of Chemistry, Michigan State University, East Lansing, MI 48824, USA}

\author{A. Stolz}
\affiliation{National Superconducting Cyclotron Laboratory, Michigan State University, East Lansing, MI 48824, USA}

\author{R. G. T. Zegers}
\affiliation{National Superconducting Cyclotron Laboratory, Michigan State University, East Lansing, MI 48824, USA}
\affiliation{Joint Institute for Nuclear Astrophysics, Michigan State University, East Lansing, MI 48824, USA}
\affiliation{Department of Physics and Astronomy, Michigan State University, East Lansing, MI 48824, USA}

\date{\today}

\begin{abstract}
The $\beta$-decay properties of the N=Z nuclei $^{96}$Cd, $^{98}$In and $^{100}$Sn have been studied. 
These nuclei were produced at the National Superconducting Cyclotron Laboratory (NSCL) by fragmenting a 120 MeV/nucleon $^{112}$Sn primary beam on a Be target.  
The resulting radioactive beam was filtered in the A1900 and the newly commissioned Radio Frequency Fragment Separator (RFFS) to achieve a purity level suitable for decay studies.  
The observed production cross sections of these nuclei are lower than predicted by factors of 10 to 30.
The $^{100}$Sn production cross section is 0.25(15) pb, in sharp contrast with the 120 pb lower limit established at 63 MeV/nucleon incident energy of the same primary beam.
The deduced half-life of $^{100}$Sn is 0.55$^{+0.70}_{-0.31}$ s, in agreement with previous measurements.
Two $\beta$-decaying states in $^{98}$In were observed with measured half-lives of 47(13) ms and 0.66(40) s.
The half-life of $^{96}$Cd, which was the last experimentally unknown waiting point half-life of the astrophysical rp-process, is 1.03$^{+0.24}_{-0.21}$ s.
The implications of the experimental T$_{1/2}$ value of $^{96}$Cd on the abundances predicted by rp-process calculations and the origin of A=96 isobars such as $^{96}$Ru are explored.
\end{abstract}


\maketitle

The region of N=Z nuclei near $^{100}$Sn has sparked interest for both nuclear structure and nuclear astrophysics during many years.  
The possibility that nearly all the Gamow-Teller (GT) strength in these nuclides lies within the $\beta$-decay Q-value window\cite{BrownPRC94} has significant impact on their role as waiting points of the astrophysical rapid proton (rp) capture process, that occurs for example in some X-ray bursts.
The rp-process is expected to reach $^{100}$Sn and end in a Sn-Sb-Te cycle in particularly powerful bursts with large amounts of hydrogen at ignition \cite{SchatzPRL01}.
Proton capture on these even N=Z nuclei is highly suppressed because the resulting isotone is either proton-unbound or destroyed by the more favored inverse reaction, due to the low Q-value for the (p,$\gamma$) reaction.
The rp-process is therefore momentarily halted and has to wait for the $\beta$-decay of the N=Z nucleus to occur.
In addition to nuclear masses, $\beta$-decay half-lives are among the most important nuclear physics data needed to predict the rp-process path, energy generation rate, and final composition.
Longer half-lives of the waiting points enhance the mass abundance at that particular isotope and shape the composition of the rp-process ashes.
Predicting this composition reliably is important for the understanding of neutron-star crusts \cite{GuptaAJ07} and possibly the origin of the nuclei $^{92,94}$Mo or $^{96,98}$Ru found in unexpectedly large quantities in the solar system \cite{WeinbergAJ06}.
$^{96}$Cd is expected to be one of these important waiting points and the direct progenitor of $^{96}$Ru in the rp-process.
Moreover, it is the last even N=Z nuclei on the rp-process path for which the half-life is still experimentally unknown.

The production and separation of $^{100}$Sn and its N=Z neighbors has proved to be a major experimental challenge.
Early attempts used heavy-ion induced fusion-evaporation reactions in combination with an on-line mass separator.
However, this method has a low luminosity due to the necessary use of thin targets, and is hindered by the presence of isobars, which are difficult to separate from the nuclei of interest.  
No positive identification of $^{100}$Sn was obtained via this method \cite{KarnyEPJA05}.  
The projectile fragmentation technique offers higher luminosity, thanks to the higher energy of the beam which enables the use of thick targets, and clear separation of the contaminants via event-by-event tagging.  
Reported observations of $^{100}$Sn using this method were made in the relativistic energy domain using 1 GeV/nucleon $^{124}$Xe and $^{112}$Sn beams \cite{SummererNPA97, FaestermannEPJA02},  as well as in the intermediate energy domain from a 60 MeV/nucleon $^{112}$Sn beam \cite{RykaczewskiPRC95}.  
While in the relativistic energy domain the level of contamination from other species in the secondary beam was acceptable for performing decay spectroscopy \cite{FaestermannEPJA02}, this is not the case at lower incident energies. 
High levels of contamination come from low momentum tails of higher rigidity fragments which extend exponentially and overlap with the momentum acceptance of the fragment separator.  
Additional filtering becomes necessary, provided for instance by a velocity filter such as the Wien filter used in the GANIL experiments \cite{LewitowiczNPA95}.  
At the NSCL, an alternative solution to this problem has recently been developed which takes advantage of the radio-frequency time structure of the beam provided by the coupled cyclotrons.  
The recently commissioned Radio Frequency Fragment Separator (RFFS) applies a transverse radio-frequency electric field to the secondary beam which deflects fragments according to their phase difference with the primary beam.  
As the phase difference itself depends on the velocity and distance travelled by the fragments, the net effect of this device is a transversal deflection according to velocity.
A slit placed downstream of the device allows a velocity filtering of the particles resulting in beams sufficiently pure for decay studies \cite{BazinNIM08}.  

\begin{figure}
\includegraphics[scale=0.5]{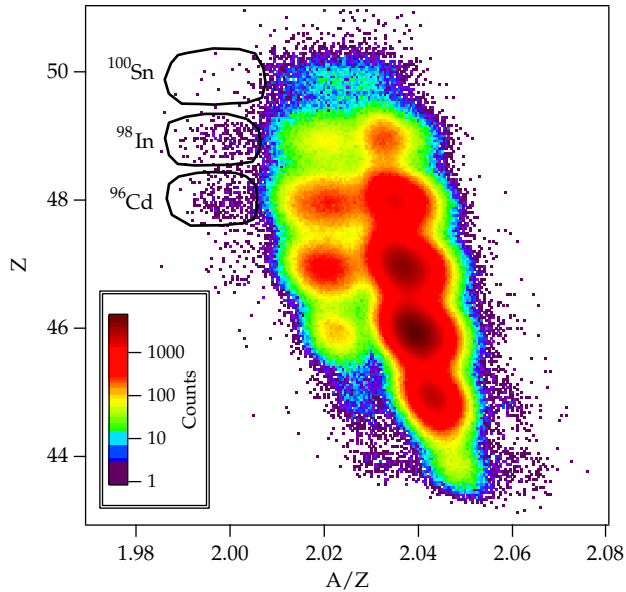}
\caption{\label{pid}(Color online) Particle identification spectrum of the heavy nuclei transmitted through the RFFS.  The most intense contaminants normally present at larger A/Z are rejected.
The low-Z contaminants are not shown on this figure.
The gates indicated on the spectrum were used to select the N=Z nuclei, in order to reduce cross contamination between elements.}
\end{figure}

A secondary beam containing the N=Z nuclei $^{96}$Cd, $^{98}$In and $^{100}$Sn was produced at the NSCL from a 120 MeV/nucleon $^{112}$Sn primary beam reacting with a 195 mg/cm$^2$ $^9$Be target placed at the target location of the A1900 Fragment Separator \cite{MorrisseyNIM03}.  
The A1900 was used as an energy-loss achromat by inserting a curve-shaped Kapton foil equivalent to a wedge profile at its dispersive focal plane.  
The thickness of the wedge was 40.6 mg/cm$^2$.  
The materials of both the target and the wedge were chosen to minimize the losses due to the production of lower charge states, and to maximize the yield of $^{100}$Sn.  
A detailed scanning of the momentum distributions of the $^{104,103,102}$Sn isotopes was performed for targets of $^9$Be and $^{58}$Ni with equivalent energy losses in order to study the effect of the reaction mechanism on the production of those rare isotopes, and extrapolate a production cross section for $^{100}$Sn.  
The $^{58}$Ni target was backed with a 47 mg/cm$^2$ thick $^9$Be foil to retain similar charge state distributions in both cases.  
This study revealed a higher yield by a constant factor of 5 in favor of the $^9$Be target for $^{104,103,102}$Sn isotopes.  
A single rigidity setting for maximum production of $^{101}$Sn revealed the same effect.  
In our experiment the maximum yields of charge states were the fully-stripped ions, which simplified the particle identification based on energy-loss and time-of-flight measurements (see Fig. \ref{pid}).
The particle identification was confirmed using the $\gamma$-ray tagging method \cite{GrzywaczPLB95} on the abundant $\mu$s isomers produced in the cocktail beam including $^{96}$Pd, $^{93}$Ru and $^{90}$Mo, which were transmitted to the experimental end station when the RFFS was turned off.

The RFFS was operated at its nominal voltage of 100 kV across the 5 cm gap of its deflector plates, while the phase was adjusted to let the N=Z heavy ions and closest neighbors pass to the experimental end station.  
Some of the low-Z contaminants, whose phase difference with the nuclei of interest was close to 2$\pi$, were also passed to the experimental end station. 
The overall contamination rate was reduced by about a factor of 200, to an average rate of about 50 counts per second, suitable for $\beta$-decay studies.  
The filtered secondary beam was implanted in the NSCL Beta Counting System (BCS) \cite{PrisciandaroNIM03}, itself surrounded by 16 detectors of the Segmented Germanium Array (SeGA) \cite{MuellerNIM01} in close geometry.  
The BCS was configured with three silicon PIN detectors in front of a 1 mm thick 40$\times$40 Double Sided Strip Detector (DSSD) in which the ions were implanted at a depth of 200--300 $\mu$m, itself followed by the $\beta$ calorimeter consisting of six 1 mm thick Single Sided Strip Detectors (SSSD) and a 1 cm thick planar high purity germanium crystal.  
The rate of the secondary beam at the dispersive image of the A1900 was too high (greater than 2 MHz) to allow for  momentum tracking with a plastic scintillator.  
The momentum acceptance was therefore limited to 1\%, the maximum value possible for separating masses close to 100 without momentum tracking.  
\begin{table}
\caption{\label{yields}Number of implantations observed for N=Z+1 and N=Z Cd, In and Sn nuclei, with their corresponding cross sections.  
The number of counts and error bars are deduced from fits of the mass spectra for each element.
The cross section error bars also take into account the systematic errors stemming from the determination of the transmission and the monitoring of the beam intensity.}
\begin{ruledtabular}
\begin{tabular}{c|c|c|c|c}
Nucleus & Counts & $\sigma_{exp}$ (pb) & $\sigma_{EPAX}$ (pb) & Ratio \\\hline
$^{97}$Cd & 1.14(1)$\times$10$^5$ & 3900(700) & 6500 & 1.7(3) \\
$^{99}$In & 3.02(9)$\times$10$^4$ & 900(200) & 1000 & 1.1(3) \\
$^{101}$Sn & 3.6(3)$\times$10$^3$ & 100(30) & 100 & 1.0(4) \\\hline
$^{96}$Cd & 274(24) & 5.5(14) & 170 & 31$^{+10}_{-6}$ \\
$^{98}$In & 216(21) & 3.8(12) & 41 & 11$^{+5}_{-3}$ \\
 $^{100}$Sn & 14(5) & 0.25(15) & 6.6 & 26$^{+40}_{-10}$ \\
\end{tabular} 
\end{ruledtabular}
\end{table}
The number of implantations observed and the corresponding cross sections are reported in Table \ref{yields}.  
A total primary beam dose of 6.7$\times$10$^{16}$ $^{112}$Sn particles was accumulated during 11.5 days, yielding an average intensity of 10.7 particle-nA. 
The transmission values due to the finite acceptance of the A1900 and the charge state distributions were calculated using the \textsc{LISE++} program \cite{TarasovNPA04}, and were between 5\% and 12\% depending on the nucleus.  
Also indicated in Table \ref{yields} are the cross sections calculated from the EPAX systematics \cite{SummererNPA02}.  
Although the systematics compare favorably with measured cross sections for N=Z+1 nuclei, the observed cross sections are between a factor of 10 and 30 below the EPAX predictions for N=Z nuclei.
Such deviation was not previously seen at relativistic energies.
For comparison, the $^{100}$Sn cross sections observed using 1 GeV/nucleon $^{112}$Sn and $^{124}$Xe primary beams were 1.8$^{+2.9}_{-1.1}$ pb and 11$^{+5.5}_{-3.7}$ pb, respectively \cite{StolzPRC02,SummererNPA97}, where the error bars have been added from the number of counts observed.  
The $^{100}$Sn cross section reported here is in sharp contrast with the 120 pb lower limit established in fragmentation of a $^{112}$Sn primary beam at 63 MeV/nucleon on a Ni target \cite{LewitowiczPLB94}.

\begin{figure}
\includegraphics[scale=0.6]{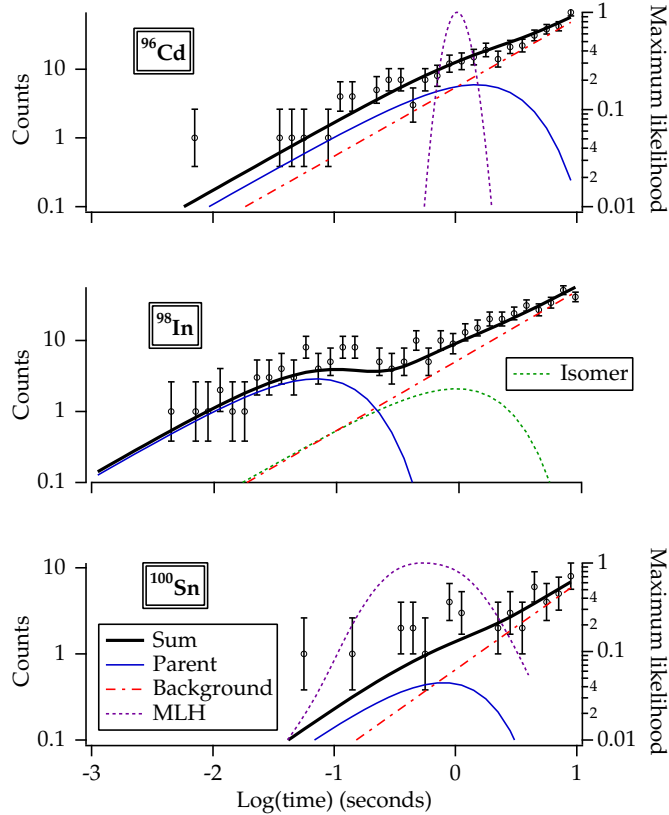}
\caption{\label{decay}
(Color online) Decay spectra and fits for the N=Z nuclei $^{96}$Cd, $^{98}$In and $^{100}$Sn displayed as logarithmic time v.s. counts.
Although included in the sum, the small contributions from the daughter and contaminant decays are not shown for simplicity.
The likelihood function for the $^{96}$Cd and $^{100}$Sn decays is also shown (scale on the right-hand side).
The additional isomeric component for the $\beta$-decay of $^{98}$In is plotted.}
\end{figure}

The $\beta$ decay of implanted nuclei was followed by tracking the decay events in the implantation pixel and its nearest neighbors for 10 seconds after each implant.
The $\beta$ efficiency of the DSSD was 37\%, obtained from fitting the decay curves of other more abundant isotopes.
Since the energy loss of the $\beta$ particles in the DSSD is a few hundred keV, it was assumed that the efficiency is the same for all Q$_\beta$ values.
A maximum likelihood method (MLH) with a predicted background has been used to minimize the error in the determination of the half lives.
This predicted  background was calculated from the implantation history in each pixel, based on the particle identification and the known half-lives of the contaminants.  
The accuracy of this prediction was checked by comparison with the actual background observed in each pixel for the N=Z+1 nuclei, which had larger statistics.
A confidence level was assigned to each event based on fits of the mass distributions for each element.
Therefore, the probability of an implantation being a true N=Z event, as opposed to a N=Z+1 contaminant, is included in the MLH analysis.  
The determination of the half-lives for the N=Z implants was then performed by running the MLH analysis while varying the minimum confidence level until an optimum was found, corresponding to the smallest error.  
In addition, the cross contamination between elements was reduced by only using implantation events selected from the gates shown in Fig. \ref{pid}.  
Both MLH and conventional fitting methods gave compatible results in the analysis of the $^{96}$Cd decay, for which the statistics is relatively large.  Since our MLH software was not tailored to handle a two-component decay, and the statistics is similar, we only used conventional fitting for the decay of $^{98}$In.
The half-lives of the two $\beta$-decaying states were determined from fitting the measured decay curve with two half-lives each including the decay of the daughter and a predicted constant background.
The $\beta$-decay spectra and fits are presented as a function of logarithmic time in Fig. \ref{decay}.  Also included in the figure are the likelihood  curves deduced from the analysis for $^{96}$Cd and $^{100}$Sn.  For $^{98}$In, the two components used in the conventional fit are also shown.

\begin{table}
\caption{\label{halflives}Measured half-lives of N=Z nuclei, compared to previous determinations and theoretical predictions.
The average between present and previous measurements is also shown, and was obtained following the prescription described in \cite{BarlowarXiv04}.}
\begin{ruledtabular}
\begin{tabular}{c|c|c|c|c}
Nucleus & $^{96}$Cd (s) & $^{98}$In$^g$ (ms) & $^{98}$In$^m$ (s) & $^{100}$Sn (s)\\\hline
This work & 1.03$^{+0.24}_{-0.21}$ & 47(13) & 0.66(40) & 0.55$^{+0.70}_{-0.31}$\\
GSI \cite{SummererNPA97} \& \cite{KienlePPNP01} & & 32$^{+32}_{-11}$ & 1.2$^{+1.2}_{-0.4}$ & 0.94$^{+0.54}_{-0.26}$\\
Average & & 44$^{+13}_{-12}$ & 0.92$^{+0.27}_{-0.17}$ & 0.86$^{+0.37}_{-0.20}$\\\hline
QRPA\cite{MoellerADNDT97} & 0.84 & 93 & & 1.81 \\
Gross Theory\cite{TakahashiADNDT73} & 0.3 & & & \\
QRPA\cite{BiehlePRC92} & 0.6 & & & \\
Shell Model\cite{HerndlNPA97} & 2.18 & 18 & & \\
Shell Model\cite{JohnstoneJPG95} & & & & 0.7 \\
Shell Model\cite{BrownPRC94} & & & & 0.53 \\
\end{tabular} 
\end{ruledtabular}
\end{table}

The half-life results are summarized in Table \ref{halflives}, together with previous measurements and theoretical predictions.  
The precision of the present result for $^{100}$Sn is similar to the previous measurement \cite{SummererNPA97}, and combining the two results yields an adopted half-life of 0.86$^{+0.37}_{-0.20}$ s.
The half-life represents an important component to determine the overall GT strength, and the next step needed for a quantitative measure of B(GT) is a the study of the branching ratios to excited states.  Such data can be inferred from the detection of $\beta$-delayed $\gamma$ rays.
Achieving this goal will require a sample size at least an order of magnitude larger than realized here.
The decay of $^{98}$In shows indication of a long-lived isomeric component and a short half-life attributed to the super-allowed Fermi decay from a 0$^+$ ground state, as previously observed in \cite{KienlePPNP01}.  
Half-lives of 44$^{+13}_{-12}$ ms and 0.92$^{+0.27}_{-0.17}$ s for the short and long components, respectively, are obtained by combining our results with the previous measurement.
The observed proportion of isomeric state populated in the fragmentation reaction is 42(20)\%, as deduced from the two component fit.
The MLH analysis for $^{96}$Cd yields a half-life value of 1.03$^{+0.24}_{-0.21}$ s, within the range of the several theoretical predictions for the ground state.  
A predicted second $\beta$-decaying state in $^{96}$Cd with a half-life of 0.5 s \cite{OgawaPRC83} could not be ruled out.

\begin{figure}
\includegraphics[scale=0.45]{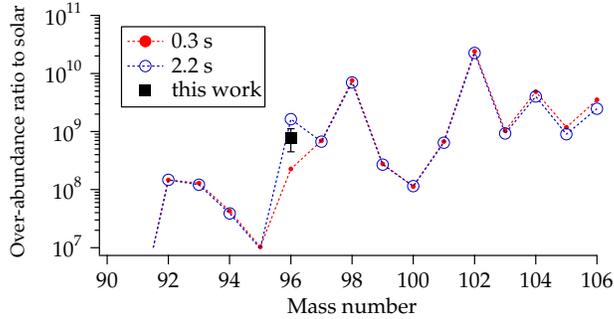}
\caption{\label{rp-process}(Color online) Predicted over-abundances relative to solar as a function of mass number, from rp-process calculations using a $^{96}$Cd half-life ranging from 0.3 s to 2.2 s, and our measured value. 
A pronounced peak at mass 96 would indicate that $^{96}$Cd plays a significant role as a waiting point.}
\end{figure}

With our half-life measurement of $^{96}$Cd, all major waiting points along the path of a strong rp-process along the proton drip line up to the Sn-Sb-Te cycle are now known experimentally. 
This allows one, for example, to reliably calculate the final composition of X-ray burst ashes with particularly high hydrogen contents at ignition \cite{SchatzPRL01}. 
To explore  the impact of our measurement on a possible explanation of the origin of $^{96}$Ru by such models, we show in Fig. \ref{rp-process} the overproduction factors (ratio of produced abundance to solar abundance) for various $^{96}$Cd half-lives. 
A rapid freeze-out of the rp-process was assumed in these calculations to conserve the A=96 masses which would be otherwise largely destroyed.
To explain the natural origin of $^{96}$Ru, a peak as high as those observed at A=98 and A=102 would be necessary.
Overproduction factors of the order of 10$^{10}$ have been suggested to be close to what is needed to explain the origin of solar $^{96}$Ru or $^{98}$Ru. Such overproduction factors could compensate for the small amount of material realistically estimated to be ejected in radius expansion X-ray bursts, and the relatively small amount of mass processed by these systems in our Galaxy \cite{WeinbergAJ06}.
With our new data this possibility is now excluded, indicating that X-ray bursts are not the main source of the large amounts of $^{96}$Ru in the solar system.

\begin{acknowledgments}
We would like to thank the operations department of the NSCL for providing the high intensity $^{112}$Sn primary beam.
This work is supported by NSF grants PHY02-16783 and PHY-06-06007.
\end{acknowledgments}

\end{document}